\documentclass[a4paper,10pt]{article}
\usepackage{amssymb}
\usepackage{amsmath}
\usepackage{epsfig}
\usepackage{subfigure}
\usepackage{graphics}
\usepackage{tikz}
\usepackage{latexsym}
\usepackage{rotating}
\usepackage{titlesec}
\usetikzlibrary{matrix}

\title{Nonlocal KdV Equations}

\author{Metin G\"{u}rses \thanks{gurses@fen.bilkent.edu.tr}\\
{\small Department of Mathematics, Faculty of Science}\\
{\small Bilkent University, 06800 Ankara - Turkey}\\
Asl{\i} Pekcan \thanks{aslipekcan@hacettepe.edu.tr} \\
{\small Department of Mathematics, Faculty of Science} \\
{\small Hacettepe University, 06800 Ankara - Turkey}
}

\date{\nonumber}
\begin{document}
\maketitle
\date{\nonumber}
\newtheorem{thm}{Theorem}[section]
\newtheorem{Le}{Lemma}[section]
\newtheorem{defi}{Definition}[section]
\newtheorem{ex}{Example}[section]
\newtheorem{pro}{Proposition}[section]
\baselineskip 17pt

\numberwithin{equation}{section}

\begin{abstract}
Writing the Hirota-Satsuma (HS) system of equations in a symmetrical form we find its local and new nonlocal reductions. It turns out that all reductions of the HS system are Korteweg-de Vries (KdV), complex KdV, and new nonlocal KdV equations. We obtain one-soliton solutions of these KdV equations by using the method of Hirota bilinearization.
\end{abstract}

\section{Introduction}

After Ablowitz and Musslimani introduced nonlocal type of reductions \cite{AbMu1}-\cite{AbMu3} we witnessed
 several works about finding new integrable nonlocal nonlinear partial differential equations and obtaining different kinds of solutions of these local and nonlocal equations. In particular, most of the works were focused on nonlocal nonlinear Schr\"{o}dinger equations (NLS) \cite{AbMu1}-\cite{jianke},
nonlocal modified Korteweg-de Vries (mKdV) equations \cite{AbMu2}-\cite{chen}, \cite{GurPek3}, \cite{GurPek2}-\cite{ma},  nonlocal sine-Gordon (SG) equations \cite{AbMu2}-\cite{chen}, \cite{aflm}, and so on \cite{fok}-\cite{hydro}.

 Recently, it has been shown that systems admitting nonlocal reductions have discrete symmetry transformations leaving the systems invariant. In \cite{origin} we showed that a special case of discrete symmetry transformations are actually the nonlocal reductions of the same systems. The connection between local and nonlocal reductions was given in \cite{Vincent}, \cite{Yang}. Among all the nonlocal equations presented so far in the literature, nonlocal KdV equation is missing. It is not possible to obtain it from the nonlocal reductions of the AKNS system. For this purpose, we use the Hirota-Satsuma (HS) system and its reductions.

The original HS system of equations \cite{HS}-\cite{GKS} is given by
\begin{align}
&ap_t=\frac{1}{2} p_{xxx}+3 p p_{x}-6 q q_{x}, \label{denk2}\\
&aq_t=-q_{xxx}-3 p q_{x}\label{denk3}.
\end{align}
The above form of the HS system is not appropriate for consistent local and nonlocal reductions. It is more useful to map it to a more symmetrical version.
By letting
\begin{equation}\label{relation}
p=\frac{1}{2}(u+v),~~~q=\gamma (u-v)
\end{equation}
with $\gamma^2=\frac{1}{4}$, we obtain \cite{DS}-\cite{Foursov}
\begin{align}
&au_t=-u_{xxx}+3v_{xxx}-6uu_x+6vu_x+12v_xu,\label{a}\\
&av_t=-v_{xxx}+3u_{xxx}-6vv_x+6uv_x+12vu_x,\label{b}
\end{align}
where $a$ is a constant.

It is possible to obtain soliton solutions of the above symmetrical HS system by the Hirota bilinearization method. The bilinearizing transformation and the Hirota bilinear form of the
original system (\ref{denk2}) and (\ref{denk3}) were given in \cite{HS}. By using the result of \cite{HS} with the transformation (\ref{relation}) and introducing two additional
constants $u_0$ and $v_0$ and by letting $u=u_0+2(\ln f)_{xx}+\frac{g}{f}$ and $v=v_0+2(\ln f)_{xx}-\frac{g}{f}$, we obtain the Hirota
bilinear form of the system (\ref{a}) and (\ref{b}) as
\begin{align}
&(aD_t+4D_x^3+(18u_0-6v_0)D_x)\{g\cdot f\}=0,\label{Hirota1}\\
&(aD_xD_t-2D_x^4-6(u_0+v_0)D_x)\{f\cdot f\}=-12g^2.\label{Hirota2}
\end{align}

In this work we find the reductions of the HS system (\ref{a}) and (\ref{b}). All the reductions of the HS system gives a kind of KdV equation. We find the standard KdV  equation
\begin{equation}
au_t=2u_{xxx}+12uu_x
\end{equation}
from the real reduction, complex KdV equation
\begin{equation}
au_t=-u_{xxx}+3\bar{u}_{xxx}-6uu_x+6\bar{u}u_x+12u\bar{u}_x
\end{equation}
from the complex reduction, nonlocal KdV equation
\begin{equation}
au_t(x,t)=-u_{xxx}(x,t)+3u_{xxx}(-x,-t)-6u(x,t)u_x(x,t)+6u(-x,-t)u_x(x,t)+12u_x(-x,-t)u(x,t)
\end{equation}
from the real nonlocal reduction, and complex nonlocal KdV equations
\begin{equation}
au_t(x,t)=-u_{xxx}(x,t)+3u_{xxx}(\varepsilon_{1} x,\varepsilon_{2} t)-6u(x,t)u_x(x,t)+6u(\varepsilon_{1} x,\varepsilon_{2} t)u_x(x,t)+12u_x(\varepsilon_{1} x,\varepsilon_{2} t)u(x,t)
\end{equation}
from the nonlocal complex reductions, where  $\varepsilon_1^2=\varepsilon_2^2=1$. We give the derivation of all these equations in Section 3.
In Section 4 we obtain the one-soliton solutions of the above local and nonlocal KdV equations by using the one-soliton solutions of the HS system obtained in Section 2 and the reduction constraints.

\section{One-soliton solutions of the HS system}

To obtain the soliton solutions of the reduced equations we need the soliton solutions of the HS system. For this purpose, we use the Hirota bilinear equations given in (\ref{Hirota1}) and (\ref{Hirota2}).

Let $g=\varepsilon g_1$ and $f=1+\varepsilon^2 f_2$ where $g_1=ce^{\theta_1}$ for $\theta_1=k_1x+\omega_1t+\delta_1$ and $c$ is a constant, in the Hirota bilinear equations
(\ref{Hirota1}) and (\ref{Hirota2}). From the coefficient of $\varepsilon$, we get the dispersion relation
\begin{equation}\label{omega_1}
\omega_1=\frac{-4k_1^3+(6v_0-18u_0)k_1}{a}
\end{equation}
if $c\neq 0$.
The coefficient of $\varepsilon^2$ gives
\begin{equation}
af_{2,xt}-2f_{2,xxxx}-6(u_0+v_0)f_{2,xx}=-6c^2e^{2k_1x+2\omega_1+2\delta_1}
\end{equation}
yielding
\begin{equation}
f_2=A_1e^{k_2x+\omega_2t+\delta_2}+A_2e^{2k_1x-\frac{8k_1^3}{a}t+2\delta_1},
\end{equation}
where
\begin{equation}\label{omega_2}
\omega_2=\frac{2k_2^3+6(u_0+v_0)k_2}{a},
\end{equation}
and
\begin{equation}\label{A_2}
A_2=\frac{c^2}{8k_1^2(k_1^2+2u_0)},
\end{equation}
and $A_1$ is an arbitrary constant. For $c\neq 0$, $A_1 \neq 0$, the coefficients of $\varepsilon^3$ and $\varepsilon^4$ vanish if
\begin{equation}\label{k_1vsk_2}
k_2=k_1\pm \sqrt{-k_1^2-4u_0}.
\end{equation}
 Take $\varepsilon=1$. Hence one-soliton solution of the
HS system (\ref{a}) and (\ref{b}) is given by the pair $(u(x,t),v(x,t))$,
{\small\begin{align}
&u(x,t)=u_0+\frac{c(e^{\theta_1}+A_2 e^{3\theta_1}+A_1e^{\theta_1+\theta_2})+8k_1^2A_2 e^{2\theta_1}+2A_1k_2^2e^{\theta_2}+2A_1A_2(2k_1-k_2)^2e^{\theta_2+2\theta_1}}{(1+A_1e^{\theta_2}+A_2 e^{2\theta_1})^2},\\
&v(x,t)=v_0+\frac{-c(e^{\theta_1}+A_2 e^{3\theta_1}+A_1e^{\theta_1+\theta_2})+8k_1^2A_2 e^{2\theta_1}+2A_1k_2^2e^{\theta_2}+2A_1A_2(2k_1-k_2)^2e^{\theta_2+2\theta_1}}{(1+A_1e^{\theta_2}+A_2 e^{2\theta_1})^2},
\end{align}}
\noindent where $\theta_1=k_1x+\omega_1t+\delta_1$, $\theta_2=k_2x+\omega_2t+\delta_2$ with the dispersion relations (\ref{omega_1}) and (\ref{omega_2}), and $A_2$ is given by (\ref{A_2}). Here $A_1, a, c, k_1, \delta_1, \delta_2$ are arbitrary constants and $k_2=k_1\pm \sqrt{-k_1^2-4u_0}$.

\section{Reductions of the HS system}

For the HS system (\ref{a}) and (\ref{b}) we have two local and two nonlocal reductions.

\vspace{0.5cm}
\noindent
{\bf (a) Local reductions}:

\noindent \textbf{(i)}\, Real reduction: $v(x,t)=ku(x,t)$, $k$ is a real constant.

The system (\ref{a}) and (\ref{b}) consistently reduces to the well-known KdV equation
\begin{equation}\label{reallocaleq}
au_t=2u_{xxx}+12uu_x.
\end{equation}

\noindent \textbf{(ii)}\, Complex reduction: $v(x,t)=k\bar{u}(x,t)$, $k$ is a real constant.

Under this reduction the system (\ref{a}) and (\ref{b}) reduces to the complex KdV equation
\begin{equation}\label{complexlocaleq}
au_t=-u_{xxx}+3\bar{u}_{xxx}-6uu_x+6\bar{u}u_x+12u\bar{u}_x,
\end{equation}
consistently, if $a=\bar{a}$.

\vspace{0.5cm}
\noindent
{\bf (b) Nonlocal reductions}:

\noindent \textbf{(i)}\, Real nonlocal reductions: $v(x,t)=ku(\varepsilon_1x,\varepsilon_2t)$, $\varepsilon_1^2=\varepsilon_2^2=1$, $k$ is a real constant.

When we apply this reduction to the system (\ref{a}) and (\ref{b}), it yields that to have a consistent reduction we must have $k=1$ and $\varepsilon_1\varepsilon_2=1$ i.e. $\varepsilon_1=\varepsilon_2=-1$. Hence we get the reduced nonlocal ST-reversal KdV equation
\begin{equation}\label{realnonlocaleq}
au_t(x,t)=-u_{xxx}(x,t)+3u_{xxx}(-x,-t)-6u(x,t)u_x(x,t)+6u(-x,-t)u_x(x,t)+12u_x(-x,-t)u(x,t).
\end{equation}

\noindent \textbf{(ii)}\, Complex nonlocal reductions: $v(x,t)=k\bar{u}(\varepsilon_1x,\varepsilon_2t)$, $\varepsilon_1^2=\varepsilon_2^2=1$, $k$ is a real constant.

This reduction reduces the system (\ref{a}) and (\ref{b}) consistently if $k=1$ and $a=\bar{a}\varepsilon_1\varepsilon_2$. Therefore we have the following nonlocal complex KdV equations:
\begin{equation}\label{complexnonlocaleq}
au_t(x,t)=-u_{xxx}(x,t)+3\bar{u}_{xxx}(\varepsilon_1x,\varepsilon_2t)-6u(x,t)u_x(x,t)+6\bar{u}(\varepsilon_1x,\varepsilon_2t)u_x(x,t)
+12\bar{u}_x(\varepsilon_1x,\varepsilon_2t)u(x,t).
\end{equation}
Explicitly, here we have three nonlocal complex KdV equations.\\

\textbf{(a)\, S-reversal nonlocal complex KdV equation:}
\begin{equation}\label{Snonlocaleq}
au_t(x,t)=-u_{xxx}(x,t)+3\bar{u}_{xxx}(-x,t)-6u(x,t)u_x(x,t)+6\bar{u}(-x,t)u_x(x,t)+12\bar{u}_x(-x,t)u(x,t),
\end{equation}
where $a$ is a pure imaginary number.\\

 \textbf{(b)\, T-reversal nonlocal complex KdV equation:}
\begin{equation}\label{Tnonlocaleq}
au_t(x,t)=-u_{xxx}(x,t)+3\bar{u}_{xxx}(x,-t)-6u(x,t)u_x(x,t)+6\bar{u}(x,-t)u_x(x,t)+12\bar{u}_x(x,-t)u(x,t),
\end{equation}
where $a$ is a pure imaginary number.\\

 \textbf{(c)\, ST-reversal nonlocal complex KdV equation:}
\begin{equation}\label{STnonlocaleq}
au_t(x,t)=-u_{xxx}(x,t)+3\bar{u}_{xxx}(-x,-t)-6u(x,t)u_x(x,t)+6\bar{u}(-x,-t)u_x(x,t)+12\bar{u}_x(-x,-t)u(x,t),
\end{equation}
where $a$ is a real number.

\section{One-soliton solutions of the reduced KdV equations}

Using the soliton solutions of the HS system obtained in Section 2 and the reduction formulas we obtain one-soliton solutions of the reduced equations.

\subsection{One-soliton solutions of the reduced local KdV equations}

\noindent \textbf{(i)}\, $v(x,t)=u(x,t)$ (KdV equation).

This reduction relation holds if $u_0=v_0$ and $c=0$ yielding $A_2=0$ and $g(x,t)=0$. Hence we obtain the one-soliton solution of the equation (\ref{reallocaleq})
as
\begin{equation}
u(x,t)=u_0+\frac{2A_1k_2^2e^{\theta_2}}{(1+A_1e^{\theta_2})^2}=u_0+\frac{k_2^2}{2}\mathrm{sech}^2\Big(\frac{\theta_2}{2}+\frac{\delta}{2}\Big),
\end{equation}
where $\theta_2=k_2x+\frac{(2k_2^3+12u_0k_2)}{a}t+\delta_2$ and $A_1=e^{\delta}$. Here $k_2, a, u_0, \delta_2, \delta$ are some arbitrary constants.
For the real parameters this solution is a soliton solution.\\

\noindent \textbf{(ii)}\, $v(x,t)=\bar{u}(x,t)$, $a=\bar{a}$ (Complex KdV equation).

Let $-k_1^2-4u_0\geq 0$ i.e. $u_0\leq -\frac{k_1^2}{4}$, $k_1, u_0\in \mathbb{R}$ yielding $k_2\in \mathbb{R}$. If we use the Type 1 approach \cite{GurPek1}, \cite{GurPek3}, \cite{GurPek2} we get the following constraints on the parameters of the one-soliton solution
of the
reduced complex local KdV equation (\ref{complexlocaleq}):
\begin{align}
&1)\, v_0=u_0,\quad c=-\bar{c},\quad k_1=\bar{k}_1,\quad e^{\delta_1}=e^{\bar{\delta}_1}, \quad e^{\delta_2}=e^{\bar{\delta}_2}, \quad A_1=\bar{A}_1,\\
&2)\, v_0=u_0,\quad c=\bar{c},\quad k_1=\bar{k}_1,\quad e^{\delta_1}=-e^{\bar{\delta}_1}, \quad e^{\delta_2}=e^{\bar{\delta}_2}, \quad A_1=\bar{A}_1,\\
&3)\, v_0=u_0,\quad c=\bar{c},\quad k_1=\bar{k}_1,\quad e^{\delta_1}=-e^{\bar{\delta}_1}, \quad e^{\delta_2}=-e^{\bar{\delta}_2}, \quad A_1=-\bar{A}_1.
\end{align}
The above sets of the constraints give similar solutions. Let us consider the first set of the constraints. Let $c=i\alpha$, $\alpha\in \mathbb{R}$.
Hence one-soliton solution of (\ref{complexlocaleq}) is
\begin{equation}
u(x,t)=u_0+\frac{i\alpha[e^{\theta_1}+A_1e^{\theta_1+\theta_2}-\frac{\alpha^2}{8k_1^2(k_1^2+2u_0)}e^{3\theta_1}]
+2A_1k_2^2e^{\theta_2}-\frac{\alpha^2}{k_1^2+2u_0}e^{2\theta_1}-\frac{A_1\alpha^2}{4k_1^2(k_1^2+2u_0)}e^{\theta_2+2\theta_1}}{\Big(1+A_1e^{\theta_2}
-\frac{\alpha^2}{8k_1^2(k_1^2+2u_0)}e^{2\theta_1}\Big)^2},
\end{equation}
or equivalently
\begin{equation}
|u(x,t)|^2=\Big(u_0+\frac{2A_1k_2^2e^{\theta_2}-\frac{\alpha^2e^{2\theta_1}}{k_1^2+2u_0}-\frac{A_1\alpha^2e^{2\theta_1+\theta_2}}{4k_1^2(k_1^2+u_0)}}
{(1+A_1e^{\theta_2}-\frac{\alpha^2e^{2\theta_1}}{8k_1^2(k_1^2+u_0)})^2 }   \Big)^2+\frac{\alpha^2\Big(e^{\theta_1}+A_1e^{\theta_1+\theta_2}-\frac{\alpha^2e^{3\theta_1}}{8k_1^2(k_1^2+u_0)}         \Big)^2}{\Big(1+A_1e^{\theta_2}-\frac{\alpha^2e^{2\theta_1}}{8k_1^2(k_1^2+u_0) } \Big)^4},
\end{equation}
where $\theta_1=k_1x+\frac{-4k_1^3-12u_0k_1}{a}t+\delta_1$, $\theta_2=k_2x+\frac{2k_2^3+12u_0k_2}{a}t+\delta_2$ with $k_2=k_1\pm \sqrt{-k_1^2-4u_0}$. For $u_0<-\frac{k_1^2}{2}\leq -\frac{k_1^2}{4}$ and
$A_1>0$, the above solution is nonsingular and bounded.\\

\noindent \textbf{Example 1.} Let us take the parameters as $k_1=\frac{1}{2}, u_0=-1, \alpha=a=A_1=1, \delta_1=\delta_2=0$. Hence one-soliton solution of (\ref{complexlocaleq}) becomes
\begin{equation}
|u(x,t)|^2=\frac{H_1(x,t)}{(1+e^{(\frac{1+\sqrt{15}}{2})x+4t}+\frac{2}{7}e^{x+11t}  )^4},
\end{equation}
where
\begin{align}
&H_1(x,t)=1+e^{x+11t}+\frac{36}{49}e^{2x+22t}+\frac{4}{49}e^{3x+33t}+\frac{16}{2401}e^{4x+44t}+\frac{8}{49}e^{(3+\sqrt{15})x+(33+3\sqrt{15})t}\nonumber\\
&+2e^{(\frac{3+\sqrt{15}}{2})x+(\frac{33-3\sqrt{15}}{2})t}+e^{(2+\sqrt{15})x+(22-3\sqrt{15})t}-(2\sqrt{15}+12)e^{(\frac{3+3\sqrt{15}}{2})x
+(\frac{33-9\sqrt{15}}{2})t}\nonumber\\
&-(2\sqrt{15}+12)e^{(\frac{1+\sqrt{15}}{2})x
+(\frac{11-2\sqrt{15}}{2})t}+(12\sqrt{15}+53)e^{(1+\sqrt{15})x+(11-3\sqrt{15})t}-(\frac{8\sqrt{15}+20}{49})e^{(\frac{5+\sqrt{15}}{2})x+(\frac{55+3\sqrt{15}}{2})t}.
\end{align}
The graph of the above solution is given in Figure 1.
\newpage
\begin{center}
\begin{figure}[h!]
\begin{minipage}{1\textwidth}
\centering
\includegraphics[angle=0,scale=.17]{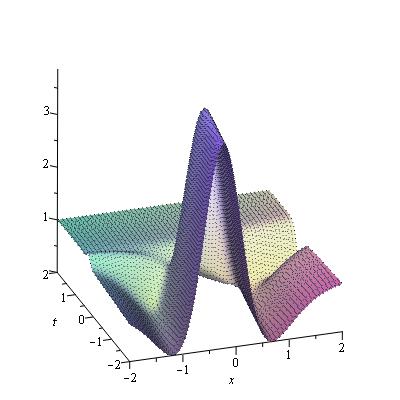}
\caption{One-soliton solution for (\ref{complexlocaleq}) with the parameters $k_1=\frac{1}{2}, u_0=-1, \alpha=a=A_1=1, \delta_1=\delta_2=0$.}
\end{minipage}
\end{figure}
\end{center}

\subsection{One-soliton solutions of the reduced nonlocal KdV equations}

\noindent \textbf{(i)}\, $v(x,t)=u(\varepsilon_1x,\varepsilon_2t)$, $\varepsilon_1^2=\varepsilon_2^2=1$ (Nonlocal KdV equation).

Recall that the system (\ref{a}) and (\ref{b}) reduces to (\ref{realnonlocaleq}) consistently  by this reduction if $\varepsilon_1=\varepsilon_2=-1$.
When we consider the reduction with the solutions $u(x,t)$ and $v(x,t)$ with Type 1 approach, we obtain that $k_1=k_2=0$ giving the trivial solution $u(x,t)=u_0$. Therefore we apply Type 2 \cite{GurPek3}, \cite{GurPek2}. In this case
we have two possibilities:
\begin{align}
 &1)\, A_1=0,\, \, u_0=v_0,\, \, e^{2\delta_1}A_2=-1,\\
 & 2)\, c=0,\, \, u_0=v_0,\, \, A_1e^{\delta_2}=\pm 1.
\end{align}
Take also all the parameters as real numbers.\\

\textbf{(i).1)}\, From the first set of the constraints we obtain the one-soliton solution of the reduced nonlocal equation (\ref{realnonlocaleq}) as
\begin{equation}
u(x,t)=u_0+\frac{c(e^{k_1x+\omega_1t+\delta_1}-e^{3k_1x+3\omega_1t+\delta_1})-8k_1^2e^{2k_1x+2\omega_1t}}{(1-e^{2k_1x+2\omega_1t})^2},
\end{equation}
where $e^{2\delta_1}=\frac{-8k_1^2(k_1^2+2u_0)}{c^2}$. This solution is a singular solution on the line $k_1x+\omega_1t=0$.\\

\textbf{(i).2)}\, In the second case we have $A_2=0$ and the one-soliton solution of the equation (\ref{realnonlocaleq}) is obtained as
\begin{equation}
u(x,t)=u_0+\sigma_1 \frac{2k_2^2e^{k_2x+\omega_2t}}{(1+\sigma_1 e^{k_2x+\omega_2t})^2},\quad \sigma_1=\pm 1.
\end{equation}
Let $\sigma_1=1$, $k_2, u_0, a \in \mathbb{R}$. Then the above solution can be rewritten as
\begin{equation}
u(x,t)=u_0+\frac{k_2^2}{2}\mathrm{sech}^2(k_2x+\omega_2t).
\end{equation}
This real-valued solution is nonsingular and bounded.\\

\noindent \textbf{Example 2.} Let us choose the parameters as $k_1=2, u_0=-1$ giving $k_2=2$, $A_1=1, \delta_2=0$ satisfying $A_1e^{\delta_2}=1$, and $a=8$. Hence one-soliton solution of (\ref{realnonlocaleq}) becomes
\begin{equation}
u(x,t)=-1+2\mathrm{sech}^2(2x-t).
\end{equation}
The graph of the above solution is given in Figure 2.
\newpage
\begin{center}
\begin{figure}[h!]
\begin{minipage}{1\textwidth}
\centering
\includegraphics[angle=0,scale=.17]{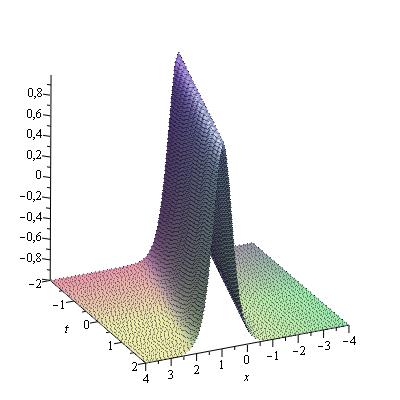}
\caption{One-soliton solution for (\ref{realnonlocaleq}) with the parameters $k_2=2, u_0=-1, a=8$.}
\end{minipage}
\end{figure}
\end{center}

\noindent \textbf{(ii)}\, $v(x,t)=\bar{u}(\varepsilon_1x,\varepsilon_2t)$, $a=\bar{a}\varepsilon_1\varepsilon_2$, $\varepsilon_1^2=\varepsilon_2^2=1$ (Nonlocal complex KdV equations).

If we use Type 1, we obtain the below constraints that must be satisfied by the parameters of the one-soliton solution of the
nonlocal reduced complex KdV equations (\ref{complexnonlocaleq}),
\begin{equation}\label{const}
k_1=\varepsilon_1\bar{k}_1, \quad v_0=\bar{u_0}=u_0, \quad a=\bar{a}\varepsilon_1\varepsilon_2
\end{equation}
with the following possibilities:
\begin{align}
&1)\, c=-\bar{c},\, A_1=\bar{A}_1,\, e^{\delta_1}=e^{\bar{\delta}_1},\, e^{\delta_2}=e^{\bar{\delta}_2},\label{nonlocalcase1}\\
&2)\, c=\bar{c},\, A_1=\bar{A}_1,\, e^{\delta_1}=-e^{\bar{\delta}_1},\, e^{\delta_2}=e^{\bar{\delta}_2},\label{nonlocalcase2}\\
&3)\, c=\bar{c},\, A_1=-\bar{A}_1,\, e^{\delta_1}=-e^{\bar{\delta}_1},\, e^{\delta_2}=-e^{\bar{\delta}_2},\label{nonlocalcase3}
\end{align}
yielding $A_2=\bar{A}_2$, $\omega_1=\varepsilon_2\bar{\omega}_1$, $\omega_2=\varepsilon_2\bar{\omega}_2$, and $k_2=\varepsilon_1\bar{k}_2$ for
\begin{equation} \label{restriction}
u_0\leq -\frac{k_1^2}{4}.
\end{equation}
Here for finding one-soliton solutions of (\ref{complexnonlocaleq}) we pick the constraints (\ref{const}) with (\ref{nonlocalcase1}). We have three types of nonlocal reduced complex equations and we will consider one-soliton solutions of them separately. Note that we will mainly focus on some particular cases giving nonsingular and bounded solutions.\\

\textbf{(a)\, S-reversal nonlocal complex KdV equation:}

In this case, according to the constraints (\ref{const}) and (\ref{nonlocalcase1}) with $a=\bar{a}\varepsilon_1\varepsilon_2$ and (\ref{restriction}), since $(\varepsilon_1,\varepsilon_2)=(-1,1)$ we have
$k_1=-\bar{k}_1$ and $a=-\bar{a}$. Let $k_1=i\alpha_1$, $k_2=i\alpha_1(1\pm \sqrt{-1+\frac{4u_0}{\alpha_1^2}})=i\alpha_2$, $c=i\alpha_3$, $a=i\alpha_4$ for
$\alpha_j, j=1,2,3,4 \in \mathbb{R}$. Hence we obtain the one-soliton solution of the S-reversal nonlocal complex equation (\ref{Snonlocaleq})
as
\begin{equation}
u(x,t)=u_0+\frac{B_1(x,t)}{(B_2(x,t))^2},
\end{equation}
where
\begin{align}
&B_1(x,t)=i\alpha_3(e^{i\alpha_1x+\omega_1t+\delta_1}+A_1e^{i(\alpha_1+\alpha_2)x+(\omega_1+\omega_2)t+\delta_1+\delta_2}+A_2e^{3i\alpha_1x+3\omega_1t+3\delta_1}             )\nonumber\\
&-2\alpha_2^2A_1e^{i\alpha_2x+\omega_2t+\delta_2}
-8\alpha_1^2A_2e^{2i\alpha_1x+2\omega_1t+2\delta_1}-2A_1A_2(2\alpha_1-\alpha_2)^2e^{i(2\alpha_1+\alpha_2)x
+(2\omega_1+\omega_2)t+2\delta_1+\delta_2}
\end{align}
and
\begin{equation}
B_2(x,t)=1+A_1e^{i\alpha_2x+\omega_2t+\delta_2}+A_2e^{2i\alpha_1x+2\omega_1t+2\delta_1},
\end{equation}
where
\begin{equation}
\omega_1=\frac{4\alpha_1^3-12u_0\alpha_1}{\alpha_4},\,\, \omega_2=\frac{-2\alpha_2^3+12u_0\alpha_2}{\alpha_4},\,\, A_2=\frac{\alpha_3^2}{8\alpha_1^2(2u_0-\alpha_1)}.
\end{equation}
If we consider the real-valued solution we get
\begin{equation}\label{Ssymmsolgeneral}
|u(x,t)|^2=\frac{H_1(x,t)}{(H_2(x,t))^2},
\end{equation}
where
\begin{align}
&H_2(x,t)=1+2A_1e^{\omega_2t+\delta_2}\cos(\alpha_2x)+2A_2e^{2\omega_1t+2\delta_1}\cos(2\alpha_1x)+A_1^2e^{2\omega_2t+2\delta_2}
+A_2^2e^{4\omega_1t+4\delta_1}\nonumber\\
&+2A_1A_2e^{(2\omega_1+\omega_2)t+2\delta_1+\delta_2}\cos((2\alpha_1-\alpha_2)x).
\end{align}
Here $H_1(x,t)$ is a huge expression so we are not giving it here explicitly. We will consider some particular cases.

\noindent \textbf{(a)} Let us consider the case when $u_0=\frac{\alpha_1^2}{3}>\frac{\alpha_1^2}{4}$ yielding $\omega_1=0$. Let also $A_1=0$. In this case we have
\begin{equation}\label{Ssymmsoln1}
|u(x,t)|^2=\frac{H_1(x,t)}{4A_2^2e^{4\delta_1}(B+\cos(2\alpha_1x))^2},\quad B=\frac{1+A_2^2e^{4\delta_1}}{2A_2e^{2\delta_1}},
\end{equation}
where
\begin{align}
&H_1(x,t)=\frac{1}{9}\alpha_1^4(A_2^4e^{8\delta_1}+1)+\alpha_3^2e^{2\delta_1}(A_2^2e^{4\delta_1}+1)+\frac{484}{9}\alpha_1^4A_2^2e^{4\delta_1}
+\frac{2}{9}\alpha_1^4A_2^2e^{4\delta_1}\cos(4\alpha_1x)
\nonumber\\
&+2A_2e^{2\delta_1}(\alpha_3^2e^{2\delta_1}-\frac{22}{9}\alpha_1^4 (A_2^2e^{4\delta_1}+1) ) \cos(2\alpha_1x)+\frac{2}{3}\alpha_1^2\alpha_3A_2e^{3\delta_1}(A_2e^{2\delta_1}-1)\sin(3\alpha_1x)
\nonumber\\
&+(\frac{2}{3}\alpha_1^2\alpha_3e^{\delta_1}(A_2^3e^{6\delta_1}-1)+\frac{44}{3}\alpha_1^2\alpha_3A_2e^{3\delta_1}(A_2e^{2\delta_1}-1)    )\sin(\alpha_1x).
\end{align}
For $B>1$ or $B<-1$ the solution (\ref{Ssymmsoln1}) is nonsingular and bounded solution.\\

\noindent \textbf{Example 3.} For this case choose the parameters as $\alpha_1=\frac{1}{2}, \alpha_3=1, \delta_1=0$. This gives $A_2=-6$. Hence one-soliton solution of (\ref{Snonlocaleq}) becomes
\begin{equation}
|u(x,t)|^2=\frac{24049+8040\cos(x)+72\cos(2x)+16968\sin(\frac{x}{2})+1008\sin(\frac{3}{2}x)}{144(37-12\cos(x))^2}.
\end{equation}
The graph of the above solution is given in Figure 3.
\begin{center}
\begin{figure}[h]
\begin{minipage}{1\textwidth}
\centering
\includegraphics[angle=0,scale=.18]{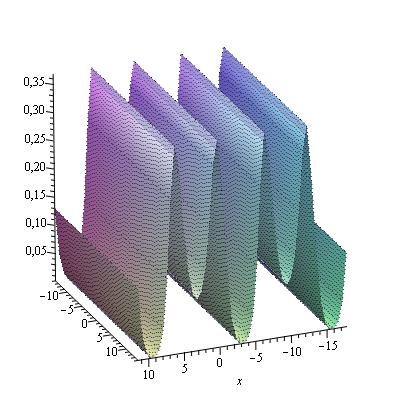}
\caption{Periodic solution for (\ref{Snonlocaleq}) with the parameters $\alpha_1=\frac{1}{2}, \alpha_3=1, \delta_1=0$.}
\end{minipage}
\end{figure}
\end{center}

\noindent \textbf{(b)} Let us assume that $\alpha_3=0$ i.e. $A_2=0$, and $u_0=\frac{\alpha_2^2}{6}$ i.e. $\omega_2=0$. Note that in this case
$\alpha_2=(3\pm \sqrt{3})\alpha_1$. Hence $u_0=\frac{\alpha_2^2}{6}>\frac{\alpha_1^2}{4}$. In this particular case, the solution (\ref{Ssymmsolgeneral})
turns to be
{\small
\begin{equation}
|u(x,t)|^2=\frac{\alpha_1^4(7\pm 4\sqrt{3})[1+A_1^4e^{4\delta_2}+100A_1^2e^{2\delta_2}-(20A_1^3e^{3\delta_2}+20A_1e^{\delta_2})\cos(\alpha_1(3\pm \sqrt{3})x)            +2A_1^2e^{2\delta_2}\cos(2\alpha_1(3\pm \sqrt{3})x)]  }{[(A_1e^{\delta_2}+\cos(\alpha_1(3\pm \sqrt{3})x))^2+\sin^2(\alpha_1(3\pm \sqrt{3})x) ]^2}.
\end{equation}}
The above solution is nonsingular and bounded if $A_1e^{\delta_2}<-1$ or $A_1e^{\delta_2}>1$.

\noindent \textbf{Example 4.} Take $\alpha_1=\frac{1}{2}, \alpha_2=\frac{1}{2}(3+\sqrt{3}), A_1=2, \delta_2=0$. Hence one-soliton solution of (\ref{Snonlocaleq}) becomes
\begin{equation}
|u(x,t)|^2=\frac{(7+4\sqrt{3})[417-200\cos(\frac{1}{2}(3+\sqrt{3})x)+8\cos((3+\sqrt{3})x)]}{16[5+4\cos(\frac{1}{2}(3+\sqrt{3})x)]^2}.
\end{equation}
The graph of the above solution is given in Figure 4.
\begin{center}
\begin{figure}[h]
\begin{minipage}{1\textwidth}
\centering
\includegraphics[angle=0,scale=.18]{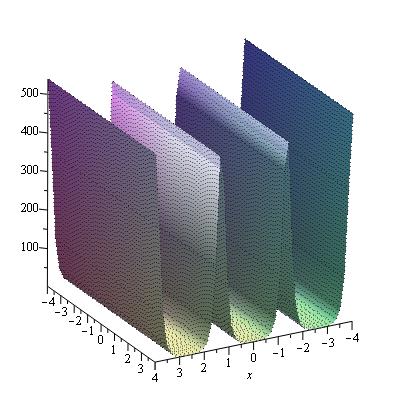}
\caption{Periodic solution for (\ref{Snonlocaleq}) with the parameters $\alpha_1=\frac{1}{2}, A_1=2, \delta_2=0$.}
\end{minipage}
\end{figure}
\end{center}

\textbf{(b)\, T-reversal nonlocal complex KdV equation:}

According to the constraints (\ref{const}) and (\ref{nonlocalcase1}) with $a=\bar{a}\varepsilon_1\varepsilon_2$ and (\ref{restriction}), since $(\varepsilon_1,\varepsilon_2)=(1,-1)$ here we have
$k_1=\bar{k}_1$ and $a=-\bar{a}$. Let $c=i\alpha_1$ and $a=i\alpha_2$ for
$\alpha_j, j=1,2, \in \mathbb{R}$ yielding
\begin{equation}
\omega_1=\frac{i(4k_1^3+12u_0k_1)}{\alpha_2}=i\alpha_3,\, \omega_2=\frac{-i(2k_1^3+12u_0k_1)}{\alpha_2}=i\alpha_4,\, A_2=-\frac{\alpha_1^2}{8k_1^2(k_1^2+2u_0)}.
\end{equation}
Therefore we obtain the one-soliton solution of the T-reversal nonlocal complex equation (\ref{Tnonlocaleq})
as
\begin{equation}
u(x,t)=u_0+\frac{B_1(x,t)}{(B_2(x,t))^2},
\end{equation}
where
\begin{align}
&B_1(x,t)=i\alpha_1(e^{k_1x+i\alpha_3t+\delta_1}+A_1e^{(k_1+k_2)x+i(\alpha_3+\alpha_4)t+\delta_1+\delta_2}+A_2e^{3k_1x+3i\alpha_3t+3\delta_1}             )\nonumber\\
&+2k_2^2A_1e^{k_2x+i\alpha_4t+\delta_2}
+8k_1^2A_2e^{2k_1x+2i\alpha_3t+2\delta_1}+2A_1A_2(2k_1-k_2)^2e^{(2k_1+k_2)x
+i(2\alpha_3+\alpha_4)t+2\delta_1+\delta_2}
\end{align}
and
\begin{equation}
B_2(x,t)=1+A_1e^{k_2x+i\alpha_4t+\delta_2}+A_2e^{2k_1x+2i\alpha_3t+2\delta_1}.
\end{equation}
Let us consider the real-valued solution $|u(x,t)|^2$. We have
\begin{equation}\label{Tsymmsolgeneral}
|u(x,t)|^2=\frac{H_1(x,t)}{(H_2(x,t))^2},
\end{equation}
where
\begin{align}
&H_2(x,t)=1+2A_2e^{2k_1x+2\delta_1}\cos(2\alpha_3t)+2A_1e^{k_2x+\delta_2}\cos(\alpha_4t)+A_1^2e^{2k_2x+2\delta_2}+A_2^2e^{4k_1x+4\delta_1}\nonumber\\
&+2A_1A_2e^{(2k_1+k_2)x+2\delta_1+\delta_2}\cos((2\alpha_3-\alpha_4)t).
\end{align}
Here again $H_1(x,t)$ is a huge expression so we are not expressing it here. Let us consider a particular case. Take $\alpha_1=0$ yielding $A_2=c=0$, and
$u_0=-\frac{k_2^2}{6}$ giving $\alpha_4=0$. Note that in this case
$k_2=(3\pm \sqrt{3})k_1$. The solution (\ref{Tsymmsolgeneral}) becomes
{\small\begin{equation}
|u(x,t)|^2=\frac{k_1^4(7\pm 4\sqrt{3})[1-20A_1e^{(3\pm \sqrt{3})k_1x+\delta_2}+102A_1^2e^{2(3\pm \sqrt{3})k_1x+2\delta_2}-20A_1^3e^{3(3\pm \sqrt{3})k_1x+3\delta_2}+A_1^4e^{4(3\pm \sqrt{3})k_1x+4\delta_2}]}{[A_1e^{(3\pm \sqrt{3})k_1x+\delta_2}+1  ]^4}.
\end{equation}}
The above solution is nonsingular and bounded for $A_1> 0$.

\noindent \textbf{Example 5.} Take the parameters as $k_1=\frac{1}{6}, k_2=\frac{1}{6}(3+\sqrt{3}), A_1=1, \delta_2=0$. Hence one-soliton solution of (\ref{Tnonlocaleq}) becomes
\begin{equation}
|u(x,t)|^2=\frac{(7+4\sqrt{3})[e^{\frac{2}{3}(3+\sqrt{3})x}-20e^{\frac{1}{2}(3+\sqrt{3})x}+102e^{\frac{1}{3}(3+\sqrt{3})x}-20e^{\frac{1}{6}(3+\sqrt{3})x}      ]}{1296(e^{\frac{1}{6}(3+\sqrt{3})x}+1)^4}.
\end{equation}
The graph of the above solution at is given in Figure 5.
\begin{center}
\begin{figure}[h]
\begin{minipage}{1\textwidth}
\centering
\includegraphics[angle=0,scale=.18]{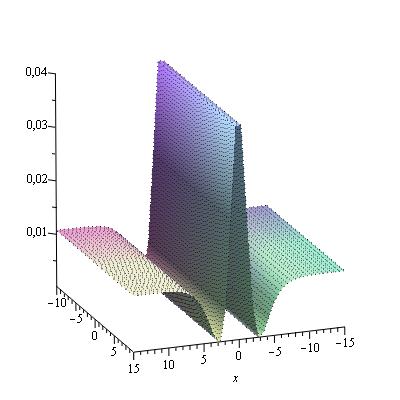}
\caption{One-soliton solution for (\ref{Tnonlocaleq}) with the parameters $k_1=\frac{1}{6}, k_2=\frac{1}{6}(3+\sqrt{3}), A_1=1, \delta_2=0$.}
\end{minipage}
\end{figure}
\end{center}

\textbf{(c)\, ST-reversal nonlocal complex KdV equation:}

In this case according to the constraints (\ref{const}) and (\ref{nonlocalcase1}) with $a=\bar{a}\varepsilon_1\varepsilon_2$ and (\ref{restriction}), since $(\varepsilon_1,\varepsilon_2)=(-1,-1)$ we have
$k_1=-\bar{k}_1$ and $a=\bar{a}$. Let $k_1=i\alpha_1$, $k_2=i\alpha_1(1\pm \sqrt{-1+\frac{4u_0}{\alpha_1^2}})=i\alpha_2$, $c=i\alpha_3$ for
$\alpha_j, j=1,2,3 \in \mathbb{R}$ yielding
\begin{equation}
\omega_1=\frac{i(4\alpha_1^3-12u_0\alpha_1)}{a}=i\alpha_4,\,\, \omega_2=\frac{i(-2\alpha_2^3+12u_0\alpha_2)}{a}=i\alpha_5,\,\, A_2=\frac{\alpha_3^2}{8\alpha_1^2(2u_0-\alpha_1^2)}.
\end{equation}
Therefore we obtain the one-soliton solution of the ST-reversal nonlocal complex equation (\ref{STnonlocaleq})
as
\begin{equation}
u(x,t)=u_0+\frac{B_1(x,t)}{(B_2(x,t))^2},
\end{equation}
where
\begin{align}
&B_1(x,t)=i\alpha_3(e^{i\alpha_1x+i\alpha_4t+\delta_1}+A_1e^{i(\alpha_1+\alpha_2)x+i(\alpha_4+\alpha_5)t+\delta_1+\delta_2}+A_2e^{3i\alpha_1x+3i\alpha_4t+3\delta_1}             )\nonumber\\
&+2A_1k_2^2e^{i\alpha_2x+i\alpha_5t+\delta_2}
+8k_1^2A_2e^{2i\alpha_1x+2i\alpha_4t+2\delta_1}-2A_1A_2(2\alpha_1-\alpha_2)^2e^{i(2\alpha_1+\alpha_2)x
+i(2\alpha_4+\alpha_5)t+2\delta_1+\delta_2}
\end{align}
and
\begin{equation}
B_2(x,t)=1+A_1e^{i\alpha_2x+i\alpha_5t+\delta_2}+A_2e^{2i\alpha_1x+2i\alpha_4t+2\delta_1}.
\end{equation}
When we consider the real-valued solution $|u(x,t)|^2$ we get
\begin{equation}\label{STsymmsolgeneral}
|u(x,t)|^2=\frac{H_1(x,t)}{(H_2(x,t))^2},
\end{equation}
where
\begin{align}
&H_2(x,t)=1+2A_2e^{2\delta_1}\cos(2\alpha_1x+2\alpha_4t)+2A_1e^{\delta_2}\cos(\alpha_2x+\alpha_5t)+A_1^2e^{2\delta_2}+A_2^2e^{4\delta_1}\nonumber\\
&+2A_1A_2e^{2\delta_1+\delta_2}\cos((2\alpha_1-\alpha_2)x+(2\alpha_4-\alpha_5)t).
\end{align}
Here also $H_1(x,t)$ is a huge expression so we are not giving it here explicitly. We will deal with some particular cases.

\noindent \textbf{(a)} Consider the case when $\alpha_3=0$ yielding $A_2=0$. In this case the solution (\ref{STsymmsolgeneral})
of the nonlocal complex equation (\ref{STnonlocaleq}) becomes
{\small
\begin{equation}\label{STa}
|u(x,t)|^2=\frac{4u_0A_1(u_0-\alpha_2^2)e^{\delta_2}(1+A_1^2e^{2\delta_2})\cos(\alpha_2x+\alpha_5t)+2u_0^2A_1^2e^{2\delta_2}\cos(2\alpha_2x+2\alpha_5t)
   +4A_1^2e^{2\delta_2}(u_0-\alpha_2^2)^2}{[(A_1e^{\delta_2}+\cos(\alpha_2x+\alpha_5t))^2+\sin^2(\alpha_2x+\alpha_5t)]^2}.
\end{equation}}
This solution is nonsingular and bounded if $A_1e^{\delta_2}>1$ or $A_1e^{\delta_2}<-1$.

\noindent \textbf{Example 6.} Take the parameters of the solution (\ref{STa}) as $\alpha_1=\frac{1}{2}$, $u_0=2$, $\delta_2=0$, $A_1=-4$, and $a=4$.
Hence the solution (\ref{STa}) turns to be
\begin{equation}
|u(x,t)|^2=\frac{1284-64(1+\sqrt{31})^2+4(1+\sqrt{31})^4+(136(1+\sqrt{31})^2-1088)\cos(\psi)+128\cos(2\psi)}{[17-8\cos(\psi)]^2},
\end{equation}
where $\psi=\frac{1}{2}(1+\sqrt{31})x+(3(1+\sqrt{31})-\frac{1}{16}(1+\sqrt{31})^3 )t$. The graph of the above solution is given in Figure 6.
\begin{center}
\begin{figure}[h]
\begin{minipage}{1\textwidth}
\centering
\includegraphics[angle=0,scale=.18]{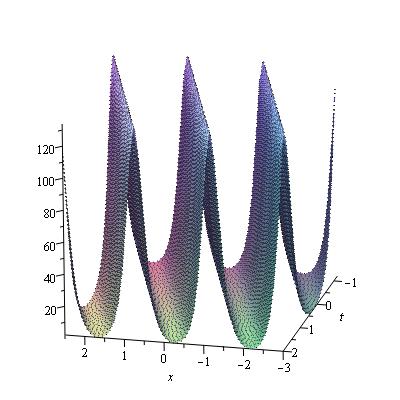}
\caption{Periodic solution for (\ref{STnonlocaleq}) with the parameters $\alpha_1=\frac{1}{2}$, $u_0=2$, $\delta_2=0$, $A_1=-4$, $a=4$.}
\end{minipage}
\end{figure}
\end{center}
\noindent \textbf{(b)} Here we again consider a particular case. Take $A_1=0$. Therefore the solution (\ref{STsymmsolgeneral})
of the nonlocal complex equation (\ref{STnonlocaleq}) becomes
\begin{equation}\label{STb}
|u(x,t)|^2=\frac{H_1(x,t)}{[(A_2e^{2\delta_1}+\cos(2\alpha_1x+2\alpha_4t))^2+\sin^2(2\alpha_1x+2\alpha_4t)]^2},
\end{equation}
where
\begin{align}
&H_1(x,t)=2\alpha_3e^{\delta_1}[2u_0A_2e^{2\delta_1}-u_0-2u_0A_2^2e^{4\delta_1}+u_0A_2^3e^{6\delta_1}-8A_2\alpha_1^2e^{2\delta_1}+8A_2^2\alpha_1^2e^{4\delta_1}   ]\sin(\alpha_1x+\alpha_4t)\nonumber\\
&+2e^{2\delta_1}A_2[2u_0^2-8u_0\alpha_1^2+\alpha_3^2e^{2\delta_1}+2u_0^2A_2^2e^{4\delta_1}-8u_0A_2^2\alpha_1^2e^{4\delta_1}]\cos(2\alpha_1x+2\alpha_4t)\nonumber\\
&+2u_0\alpha_3A_2e^{3\delta_1}[A_2e^{2\delta_1}-1]\sin(3\alpha_1x+3\alpha_4t)+2u_0^2A_2^2e^{4\delta_1}\cos(4\alpha_1x+4\alpha_4t)+u_0^2\nonumber\\
&+e^{2\delta_1}[\alpha_3^2+64A_2^2\alpha_1^4e^{2\delta_1}-32u_0A_2^2\alpha_1^2e^{2\delta_1}+4u_0^2A_2^2e^{2\delta_1}+\alpha_3^2A_2^2e^{4\delta_1}+u_0^2A_2^4e^{6\delta_1}].
\end{align}
This solution is nonsingular and bounded if $A_2e^{2\delta_1}>1$ or $A_2e^{2\delta_1}<-1$.

\noindent \textbf{Example 7.} Choose the parameters of the solution (\ref{STb}) as $\alpha_1=\frac{1}{2}, u_0=2, \delta_1=0, \alpha_3=5, a=4$. Therefore
the solution (\ref{STb}) becomes
{\small
\begin{equation}
|u(x,t)|^2=\frac{68449+45780\sin(\frac{1}{2}x-\frac{23}{8}t)+12600\sin(\frac{3}{2}x-\frac{69}{8}t)+39660\cos(x-\frac{23}{4}t)+7200\cos(2x-\frac{23}{2}t)}
{(60\cos(x-\frac{23}{4}t)+109)^2}.
\end{equation}}
The graph of the above solution is given in Figure 7.
\begin{center}
\begin{figure}[h]
\begin{minipage}{1\textwidth}
\centering
\includegraphics[angle=0,scale=.18]{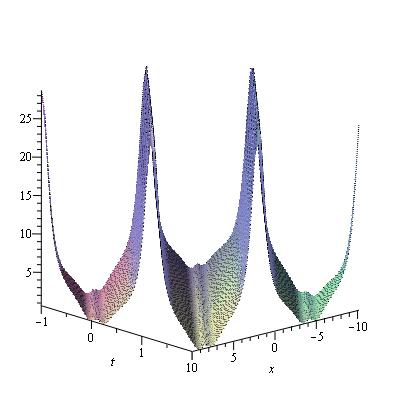}
\caption{Periodic solution for (\ref{STnonlocaleq}) with the parameters $\alpha_1=\frac{1}{2}, u_0=2, \delta_1=0, \alpha_3=5, a=4$.}
\end{minipage}
\end{figure}
\end{center}

\section{Conclusion}
In this work, we have studied local and nonlocal reductions of the symmetrical HS system. The local reductions yield KdV and complex KdV equations.
The nonlocal reductions of the HS system give several type of new nonlocal KdV equations. These are time (T)-, space (S)-, and space-time (ST)-reversal nonlocal complex KdV equations and space-time (ST)-reversal nonlocal KdV equation. By using the one-soliton solution of the HS system and the reductions formulas we obtained the one-soliton solutions of the reduced local and nonlocal KdV equations. Using our approach one can find also two and three soliton solutions of these KdV equations.

\section{Acknowledgment}
  This work is partially supported by the Scientific
and Technological Research Council of Turkey (T\"{U}B\.{I}TAK).

\end{document}